\RequirePackage[mathlines]{lineno}
\documentclass[prd,twocolumn,showpacs,amsmath,amssymb]{revtex4-1}

\usepackage{overpic,graphicx}
\usepackage{dcolumn}
\usepackage{bm}
\usepackage{rotating}
\usepackage{subfigure}
\usepackage{color}

\usepackage{slashed}
\usepackage{stfloats}
\usepackage{amsmath}
\definecolor{gongnvlan}{rgb}{0.15,0.39,0.46}
\begin{document}

\title{\bf \boldmath
    Polarization difference between hyperons and anti-hyperons induced by external
    magnetic field
}
\author{Hai-Bo Li$^{1,2}$} \email{lihb@ihep.ac.cn}

\author{Xin-Xin Ma$^{1,2}$} \email{maxx@ihep.ac.cn}

\affiliation{
    $^{1}$Institute of High Energy Physics, Beijing 100049, People's Republic of China
}
\affiliation{
    $^{2}$University of Chinese Academy of Sciences,  Beijing 100049, People's Republic of China
}

\date{\today}

\begin{abstract}
    We investigate the quantum correlated  $\Lambda \bar{\Lambda} $ production
    in the reaction
    $e^{+}e^{-} \to J/\psi \to \Lambda \bar{\Lambda}$. Since the $\Lambda$ or
    $\bar{\Lambda}$ has a nonzero magnetic moment,  its spin will undergo a
    Larmor precession in the magnetic field of the detector, such as the BESIII
    experiment. Because of the spin precession, the angular distribution of the
    $\Lambda$ and $\bar{\Lambda}$ is slightly modified.
    Therefore, we  obtain the corresponding term of the modified angular
    distribution due to the effect of the Larmor precession.
    We also estimate its potential effect on the measurements of $CP$ violation,
    as well as the decay asymmetry parameter and polarization of $\Lambda$. 
    The polarization of the $\Lambda$ or $\bar{\Lambda}$  at the production vertex will rotate around the $B$-field axis, over an angle depending on the flight length in, but it still could be measured by fit to the corrected angular distribution. 
    Of important, We conclude that a nonzero $CP$ asymmetry of order $10^{-4}$
    will be caused once neglecting spin precession of the $\Lambda$ and
    $\bar{\Lambda}$ in the  $e^{+}e^{-} \to J/\psi \to \Lambda \bar{\Lambda}$
    process.
    The size of this $CP$ asymmetry is several times  that of predicted within
    Standard Model in the hyperon decay.
    Although this effect is small, it will play an important role in future high precision experiments, such as the super-tau-charm factory.
\end{abstract}


\maketitle
\oddsidemargin  -0.2cm
\evensidemargin -0.2cm

\section{Introduction}
Since the inner parts of proton found~\cite{Chambers:1956zz},  probing the structure of baryons is still
active. However, a complete observation of the electromagnetic (e.m.) structure
of hadrons is possible merely in polarization experiments. The results for elastic scattering were firstly presented by the SLAC scattering experiments~\cite{Alguard:1976bk}, in which both electrons and proton target are polarized.
The electron-positron collider provides
coherent hyperon-anti-hyperon pairs.
In 2019,  the BESIII experiment has collected about  $10^{10}$ $J/\psi$ decay events,  which is an ideal place to
probe the form factors  and search for the $CP$ violation in the coherent $\Lambda \bar{\Lambda}$
pair production~\cite{Li:2016tlt}.
Recently, the most precision asymmetry parameter of $\Lambda$ is
measured to be $0.750 \pm 0.009 \pm 0.004$~\cite{Ablikim:2018zay}, with more
than 7.0 $\sigma$ deviation from previous world averaged value~\cite{PDG}.
The BESIII detector consists mainly of a cylindrical main draft chamber,
with a magnetic field of 1.0 T parallel to the electron beam \cite{Li:2016tlt}.
The $\Lambda$ and $\bar{\Lambda}$ are produced in the $e^+ e^-$ collision point,
due to the long lifetime of hyperon,  they will decay in flight and the average
decay length or flight length will be 12 cm in the $B$-field (1.0 T) of the  BESIII
detector.   Therefore, the hyperon will undergo a Larmor precession in the
magnetic field in the detector.  However, this effect was not considered in
previous publications \cite{Ablikim:2018zay, Ablikim:2017tys,
Ablikim:2005cda,Aubert:2007uf, Ablikim:2012bw, Aubert:2009am}.
Although this effect is small but maybe not negligible in the measurements of
$CP$ violation in hyperon decay at future high precision experiments,  such as
the proposed super-tau-charm factory~\cite{Bigi:2017eni}, in which the
sensitivities on the $CP$ measurements will reach $10^{-4}$ or even
$10^{-5}$~\cite{Bigi:2017eni}, while the $A_{CP}$ predicted by Standard-Model
(SM) is order $10^{-5}$ \cite{Donoghue:1986hh, Tandean:2002vy}.
In this case,  one has to consider the spin
precession effect which will modify the angular distribution of the $e^+ e^- \to
J/\psi \to \Lambda \bar{\Lambda}$ process,  therefore the CP asymmetry $A_{CP} =
\frac{\alpha_{-}  + \alpha_{+}}{\alpha_{-}  - \alpha_{+}}$ will be biased, and
nonzero $A_{CP}$ will be extracted once neglecting the Larmor precession. 
The paper is divided into
two parts. In the first part, we will consider this effect and derive the
corresponding modification on overall angular distributions of the final
states.
In the second part, we perform a Monte-Carlo (MC) Simulation and then give the impact
on the measurements of $\alpha_{\pm}$ and the $CP$ asymmetry parameters.

\section{The production  of $\Lambda\bar{\Lambda}$ pairs}

The coherent $\Lambda \bar{\Lambda}$ pairs are produced via the process $e^+e^- \to J/\psi \to \Lambda \bar{\Lambda}$.
The $\Lambda$ and $\bar{\Lambda}$  with intrinsic magnetic moment will undergo a
Larmor precession in the external magnetic field of the detector,  so the spin direction will be changed in the flight before its decaying.   This effect will
modify the angular distribution of the process  $e^+e^- \to J/\psi \to \Lambda
\bar{\Lambda}$.
The effective amplitude for $e^+e^- \to J/\psi \to \Lambda \bar{\Lambda}$ can be
written as
\begin{equation}
    \begin{split}
        M &= \frac{ie^2}{q^{2}} j_{\mu} \bar{u}(p_{1}, s_1) \left( F^H_1(q^2) \gamma_{\mu} +
        \frac{F^H_2(q^2)}{2m_{\Lambda}} p_{\nu} \sigma^{\nu\mu}  \gamma_5 \right) \\
        & \times \nu(p_2, s_2),
    \end{split}
\end{equation}
where $j_{\mu} = \bar{u}(k_1) \gamma_{\mu} \nu(k_2)$ is the lepton current with
$k_1$ and $k_2$ the momenta of $e^{-}$ and $e^{+}$, $m_{\Lambda}$ the mass of $\Lambda$, $s=p^2$
$q = p_1 + p_2$ and $p = p_1 - p_2$ with the $p_1$ and $p_2$ the momenta of $\Lambda$ and
$\bar{\Lambda}$,  and $s_1$ and $s_2$ the spin four-vectors of $\Lambda$ and
$\bar{\Lambda}$.  Of
important, the form factors $G^H_{E}$ and $G^H_{M}$ are usually called as
hadronic form
factors~\cite{Faldt:2016qee}, because the $\Lambda\bar{\Lambda}$ are produced
via the $J/\psi$ hadronic decay~\cite{Faldt:2017kgy},  are related to
$F^H_{1,2}$ by
\begin{equation}
    G^H_{M} = F^H_1 + F^H_2, ~  G^H_{E} = F^H_1 + \tau F^H_2,
\end{equation}
with $\tau = \frac{q^2}{4 m_{\Lambda}^2}$.
Following the method in Refs. \cite{Dubnickova:1992ii,Gakh:2005hh,Czyz:2007wi}, the differential cross section takes the following form with all
constants dropped
\begin{equation}
    \begin{split}
        \frac{d\sigma}{d \cos\theta}  \sim & 1+ \alpha _{\psi } \cos^2\theta
        + \sin ^2\theta \hat{s}_{1}^x \hat{s}_{2}^x
        + \alpha_{\psi} \sin^2\theta \hat{s}_{1}^y \hat{s}_{2}^y \\
        &- \left( \alpha_{\psi} +\cos^2 \theta  \right)\hat{s}_{1}^z \hat{s}_{2}^z
        + \sqrt{1-\alpha_{\psi}^2} \cos\Phi  \sin\theta \\
        & \times    \cos\theta
        \left( \hat{s}_{1}^x \hat{s}_{2}^z  - \hat{s}_{1}^z \hat{s}_{2}^x\right)
        +\sqrt{1-\alpha _{\psi}^2}
        \sin\Phi  \sin\theta \\
        & \times \cos\theta  (\hat{s}_{1}^y - \hat{s}_{2}^y),
    \end{split}
    \label{eq:raw}
\end{equation}
where $\theta$ is the angle between momenta of $e^{+}$ and $\Lambda$,
$\alpha_{\psi} = \frac{\tau |G_{M}|^{2} -  |G_{E}|^{2}}{\tau |G_{M}|^{2} + |G_{E}|^{2}}$,
$\hat{s}^{i}_1$ ($\hat{s}^{i}_2$) the $i$-th component of the unit vector
pointing to the direction of the $\Lambda$ ($\bar{\Lambda}$) spin in the rest
frame of its mother particle with the Z-axis direction defined by
$\Lambda$ ($\bar{\Lambda}$) momentum direction and Y-axis direction defined by
$\vec{p}_1 \times \vec{k}_2$ ($\vec{p}_2 \times \vec{k}_2$ ). $\Phi$ is the relative
phase of form factors,  
\begin{equation}
    \frac{G^H_{E}}{G^H_{M}} = e^{i \Phi} \left|\frac{G^H_{E}}{G^H_{M}} \right |.
\end{equation}

\section{$\Lambda$ spin precession}

Considering the interaction between the $\Lambda$ and the external magnetic field of the BESIII detector, we can easily
obtain \cite{Sakurai:2011zz}
\begin{equation}
    \hat{s}'_{1} = \hat{s}_{1} + \omega \tau_{\Lambda} \hat{B}  \times
    \hat{s}_{1},
\end{equation}
where $\hat{s}'_{1}$ denote the spin of $\Lambda$ in its rest frame at decay
time $\tau_{\Lambda}$
since  produced, $\hat{B}$ the direction of magnetic field in the $\Lambda$ rest
frame,
$\omega$ the precession frequency which depends
on the magnetic field magnitude $B$  and the magnetic moment of $\Lambda$, can
be written as
\begin{equation}
    \omega  = - \frac{2 \mu_{\Lambda} B}{\hbar},
\end{equation}
where the $\mu_{\Lambda}$ is the magnetic moment with the world average value
$-0.613 \pm 0.04 \mu_{N}$~\cite{PDG}. If one takes $B=1T$, lifetime of $\Lambda$
$\tau_{\Lambda} = 2.632 \times 10^{-10}$ s, and the momentum of $\Lambda$  is about 1
GeV/$c$ in the rest frame of $J/\psi$,  the average
precession angle can be determined to be about $\mathcal{A}_{\rm rota} = \omega
\tau_{\Lambda} =0.017$ rad, which will potentially contribute to the decay
parameters measurement. Similarly, for
other hyperons, $\Xi$, $\Omega$, $\Sigma^{\pm}$, the spin precession should also be
considered.

After considering this effect, the spin of $\Lambda$ became $\hat{s}'_{1}$ when it
decays in flight, so the decay amplitude of $\Lambda \to p \pi^-$ could be written as
\begin{equation}
    |M_{1}|^{2} \sim 1 + \alpha_{-} \hat{s}'_{1} \cdot n_{p},
\end{equation}
where $\alpha_{-}$ is so called decay parameter of $\Lambda$, as well as $\alpha_{+}$ the
decay parameter for $\bar{\Lambda}$,
$n_p$ the flight direction of proton in the hyperon rest-frame.
Usually the $CP$ asymmetry is defined as $A_{CP} = \frac{\alpha_{-} +
    \alpha_{+}}{\alpha_{-}- \alpha_{+}}$. Recently, the $A_{CP}$ is measured to be
$A_{CP} = -0.006 \pm 0.012 \pm 0.007$ \cite{Ablikim:2018zay},  while the theoretical
prediction within the SM is order $10^{-5}$~\cite{Donoghue:1986hh, Tandean:2002vy}.

\begin{figure}[!h]
    \centering
    \vspace{1cm}
    \includegraphics[width=0.8\linewidth]{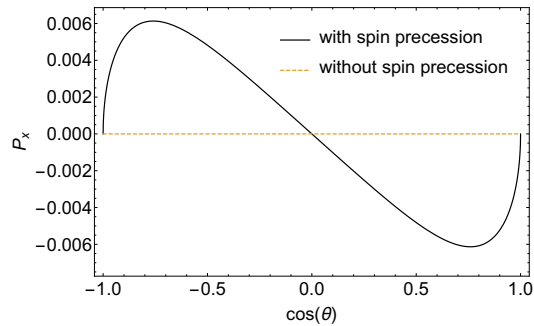}
    \caption{
        The solid black and dashed orange lines denote the $P_{\Lambda}^{y}$ of
        $\Lambda$ with and without spin precession effect, respectively.
    }
    \label{fig:px}
\end{figure}

After undergoing a spin precession in the magnetic field $\vec{B}$,  at the decay
time $\tau_{\Lambda}$, the spin of
$\Lambda$ becomes
\begin{equation}
    \left(
    \begin{array} {c}
        \hat{s}'^x_{1} \\
        \hat{s}'^y_{1} \\
        \hat{s}'^z_{1} \\
    \end{array}\right)
    =
    \left(
    \begin{array}{ccc}
        1                       & - \mathcal{A}_{\rm rota} \hat{B}'_{z}  & \mathcal{A}_{\rm rota} \hat{B}'_{y}  \\
        \mathcal{A}_{\rm rota}  \hat{B}'_{z}  & 1                       & -\mathcal{A}_{\rm rota}\hat{B}'_{x} \\
        -\mathcal{A}_{\rm rota} \hat{B}'_{y} & \mathcal{A}_{\rm rota}   \hat{B}'_{x} & 1                      \\
    \end{array}
    \right)
    \left(
    \begin{array} {c}
        \hat{s}^x_{1} \\
        \hat{s}^y_{1} \\
        \hat{s}^z_{1} \\
    \end{array}\right),
\end{equation}
\\ \\
where $\hat{B}'=\hat{B} +
(\gamma-1) (\hat{B} \cdot n_{\Lambda}) n_{\Lambda}  $ with $n_{\Lambda}$ the
flight direction of $\Lambda$ in the rest frame of $J/\psi$~\cite{landau1952the}.
Then we will average the spin of $\Lambda$, and apply the relationship
\begin{equation}
    <\hat{s}^{i} \hat{s}^{j} > = \delta^{ij}
\end{equation}
Then we will obtain the total differential cross section for the full decay chain, in which
the $\Lambda$ ($\bar{\Lambda}$) decay into $p\pi^{-}$ ($\bar{p}\pi^+$).
Here what we need to do is to replace $(s_1^{x}, s_1^{y}, s_1^{z})$ for $\Lambda$ with
\begin{equation}
    \left(
    \begin{array}{ccc}
        1                        & \mathcal{A}_{\rm rota} \hat{B}'_{z}     & -\mathcal{A}_{\rm rota} \hat{B}'_{y} \\
        - \mathcal{A}_{\rm rota}  \hat{B}'_{z} & 1                         
        & \mathcal{A}_{\rm rota} \hat{B}'_{x}  \\
        \mathcal{A}_{\rm rota} \hat{B}'_{y}    & - \mathcal{A}_{\rm rota}  \hat{B}'_{x} & 1                      
    \end{array}
    \right)
    \left(
    \begin{array} {c}
        \alpha_{-} n^x _{p} \\
        \alpha_{-}  n^y_{p} \\
        \alpha_{-}  n^z_{p} 
    \end{array}
    \right) ,
\end{equation}
so as $s_2$ for $\bar{\Lambda}$. Then the differential cross section can be obtained as
where the $\Omega_{1,2}$ is the solid angle of proton and anti-proton in the

\begin{equation}
    \label{eq:pdf}
    \begin{split}
        &\frac{d\sigma}{d \cos\theta d\Omega_{1} d\Omega_{2}}
        \sim  1+ \alpha _{\psi } \cos^2\theta
        + \sin ^2\theta \alpha_- \alpha_+ n_{p}^{x} n_{\bar{p}}^x
        + \alpha_{\psi}  \alpha_{-}  \\
        & \times \alpha_{+} \sin^2\theta
        n_{p}^y n_{\bar{p}}^y 
        - \left( \alpha_{\psi} +\cos^2 \theta  \right) \alpha_- \alpha_+ 
        n_{p}^z n_{\bar{p}}^z
        + \sqrt{1-\alpha_{\psi}^2}  \\
        & \times \alpha_- \alpha_+  \cos\Phi  \sin\theta
        \cos\theta 
        \left( n_{p}^x n_{\bar{p}}^z  - n_{p}^z n_{\bar{p}}^x\right)
        +\sqrt{1-\alpha _{\psi}^2} \sin\Phi\\ 
        & \times   \sin\theta \cos\theta  
        (\alpha_- n_{p}^y - \alpha_+ n_{\bar{p}}^y)
        + \alpha _- \alpha _+ \mathcal{A}_{\rm rota}  \sin ^2\theta
        \left(\hat{B}'_z \left(\hat{n}_{\bar{p}}^x
        \hat{n}_{p}^y \right. \right.  \\
        & \left. \left. -\hat{n}_{p}^x \hat{n}_{\bar{p}}^y\right)-\hat{B}'_y
        \left(\hat{n}_{p}^x \hat{n}_{\bar{p}}^z+\hat{n}_{\bar{p}}^x
        \hat{n}_{p}^z\right)\right)
        + \alpha_{\psi}
        \alpha _- \alpha _+ \mathcal{A}_{\rm rota} \sin^2\theta 
        \\ & \times \left(-\hat{B}'_z \hat{n}_{p}^x
        \hat{n}_{\bar{p}}^y+\hat{B}'_z \hat{n}_{\bar{p}}^x \hat{n}_{p}^y-\hat{B}'_x
        \hat{n}_{p}^y \hat{n}_{\bar{p}}^z+\hat{B}'_x \hat{n}_{\bar{p}}^y
        \hat{n}_{p}^z\right)
        - \alpha _+  \alpha _- \mathcal{A}_{\rm rota} 
        \\ & \times   \left( \alpha_{\psi} +\cos^2 \theta  \right)
        \big(\hat{B}'_y \hat{n}_{p}^x
        \hat{n}_{\bar{p}}^z
        -\hat{B}'_x \hat{n}_{p}^y \hat{n}_{\bar{p}}^z
        +\hat{B}'_y
        \hat{n}_{\bar{p}}^x \hat{n}_{p}^z+\hat{B}'_x \hat{n}_{\bar{p}}^y
        \hat{n}_{p}^z \big)
        \\ & + \mathcal{A}_{\rm rota}  \alpha _- \alpha _+ \sqrt{1-\alpha_{\psi}^2} \cos\Phi  \sin\theta
        \cos\theta \Big(\hat{B}'_x \hat{n}_{p}^x
        \hat{n}_{\bar{p}}^y+\hat{B}'_x \hat{n}_{\bar{p}}^x \hat{n}_{p}^y
        \\ &
        +\hat{B}'_z
        \hat{n}_{p}^y \hat{n}_{\bar{p}}^z
        +\hat{B}'_z \hat{n}_{\bar{p}}^y
        \hat{n}_{p}^z\Big)
        + \mathcal{A}_{\rm rota} \sqrt{1-\alpha _{\psi}^2}
        \sin\Phi  \sin\theta \cos\theta    \\
        & \times \left(
        \alpha_+ \hat{B}'_x \hat{n}_{\bar{p}}^z
        -\alpha _- \hat{B}'_z \hat{n}_{p}^x+\alpha _-
        \hat{B}'_x \hat{n}_{p}^z
        - \alpha _+ \hat{B}'_z \hat{n}_{\bar{p}}^x
        \right)
        ,
    \end{split}
\end{equation}
rest frame of $\Lambda$ and $\bar{\Lambda}$, respectively.
We should notice that the spin precession could modify the $\Lambda$ polarization state
\begin{equation}
    \begin{aligned}
        P_{\Lambda}^{x} & = - \mathcal{A}_{\rm rota} \hat{B}'_{z} P_{\Lambda}^{y} \\
        P_{\Lambda}^{z} & =  \mathcal{A}_{\rm rota} \hat{B}'_{x}
        P_{\Lambda}^{y},  
    \end{aligned}
\end{equation}
where the $P_{\Lambda}^{x,y,z}$ denote the polarization projection on the $x$,
$y$ and $z$ axis. The $P_{\Lambda}^{x,z}$ must be zero if no spin
precession, as shown in Fig. \ref{fig:px}.

\section{Monte Carlo simulation and results}

Because the spin precession is usually neglected or missed in the current experimental studies, the MC
simulation is essential for numerical study on the effect of spin precession.
The parameters $\alpha_{\psi}$, $\Phi$ and $\alpha_{\pm}$ are set
according to the measurement result in Ref.~\cite{Ablikim:2018zay}. Exactly, we
take $\alpha_{\psi}=0.462$, $\Phi=0.738$ and $\alpha_{\pm}=\pm0.750$ assuming no
$CP$ violation.
In the BESIII experiment, the magnitude of the magnetic field is around 1T.  The lifetime of the $\Lambda$ that depends on its
momentum also strongly affect the precession angle, here we take the momentum of
$\Lambda$ at $\sqrt{s/4 - m^2}$ with $\sqrt{s}$ collider energy 3.097 GeV.

The MC simulation is performed based on ROOT \cite{ROOT}. Firstly the phase space
events are generated, then an acceptance-rejection method is adopted to get the
signal toy MC samples based on the distribution in Eq. (\ref{eq:pdf}).

To reveal the effect on the measurement of the parameters $\alpha_{\psi}$,
$\Phi$ and $\alpha_{\pm}$, especially the $A_{CP}$, we perform the maximum
likelihood fit to the toy MC samples.

The probability distribution function is defined  as
\begin{equation}
    \mathcal{P} = \frac{1}{N} \frac{{\rm d} \sigma}
    {{\rm d } \cos\theta {\rm d} \Omega_1 {\rm d} \Omega_2},
\end{equation}
where $N$ is the normalization factor which is determined to
be $(4\pi)^{2}(1 + \alpha_{\psi}/3)$. The likelihood is defined as
\begin{equation}
    - \ln\mathcal{L} = -\sum_{i=1}^{n} \ln \mathcal{P}_{i},
\end{equation}
where $i$ denotes the $i$-th events in the MC sample, $n$ is the total number of
events in the MC sample which is set at $1 \times 10^{6}$.
The fitted value of the parameter with Eq. \ref{eq:pdf} is defined as  $\alpha_{\pm}^{\rm truth}$.
Then we remove this effect, in which the precession frequency is just fixed at
zero so that the Eq. \ref{eq:pdf} will be same as Eq. \ref{eq:raw}, then fit to
the same toy MC sample again,
the fitted value of decay parameters is referred as $\alpha_{\pm}^{\rm biased}$.
The differences between the results of the two fits are
\begin{equation}
    \begin{aligned}
        \Delta \alpha_{\pm} & = \alpha_{\pm}^{\rm biased} - \alpha_{\pm}^{\rm truth} 
        .
    \end{aligned}
\end{equation}

\begin{figure}[h!]
    \centering
    \includegraphics[width=0.8\linewidth]{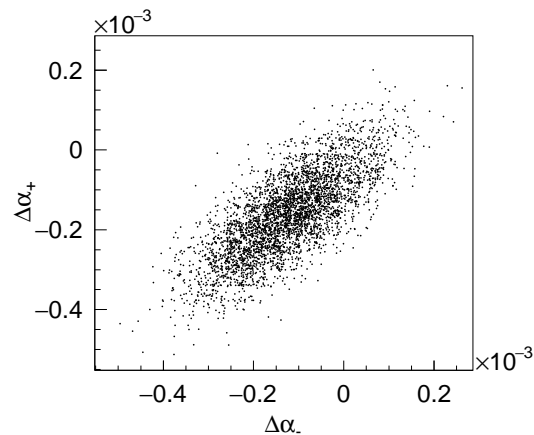}
    \caption{The strongly correlation between $\Delta \alpha_{-}$ and $\Delta \alpha_{+}$. Each black point denotes the result from fitting to each toy MC sample.}
    \label{fig:corr}
\end{figure}
We generate 4000 toy MC samples with the same data size, and find that the strong
correlation between $\Delta \alpha_{-}$ and $\Delta \alpha_{+}$ as shown in Fig.
\ref{fig:corr}.
This strong correlation leads to 
\begin{equation}
    \begin{split}
       \Delta A_{CP} &= \frac{1}{n}  \sum_{i=1}^{n} 
         \frac{\Delta \alpha_{-, i} + \Delta \alpha_{+, i}}
         {\alpha_{-, i}^{\rm biased} - \alpha_{+, i}^{\rm biased}} \\
        &=  (-1.9 \pm 0.1) \times 10^{-4}, 
    \end{split}
\end{equation}
\begin{figure}[htpb]
    \centering
    \subfigure[]{
        \label{fig:acp}
        ~~ ~\includegraphics[width=0.8\linewidth]{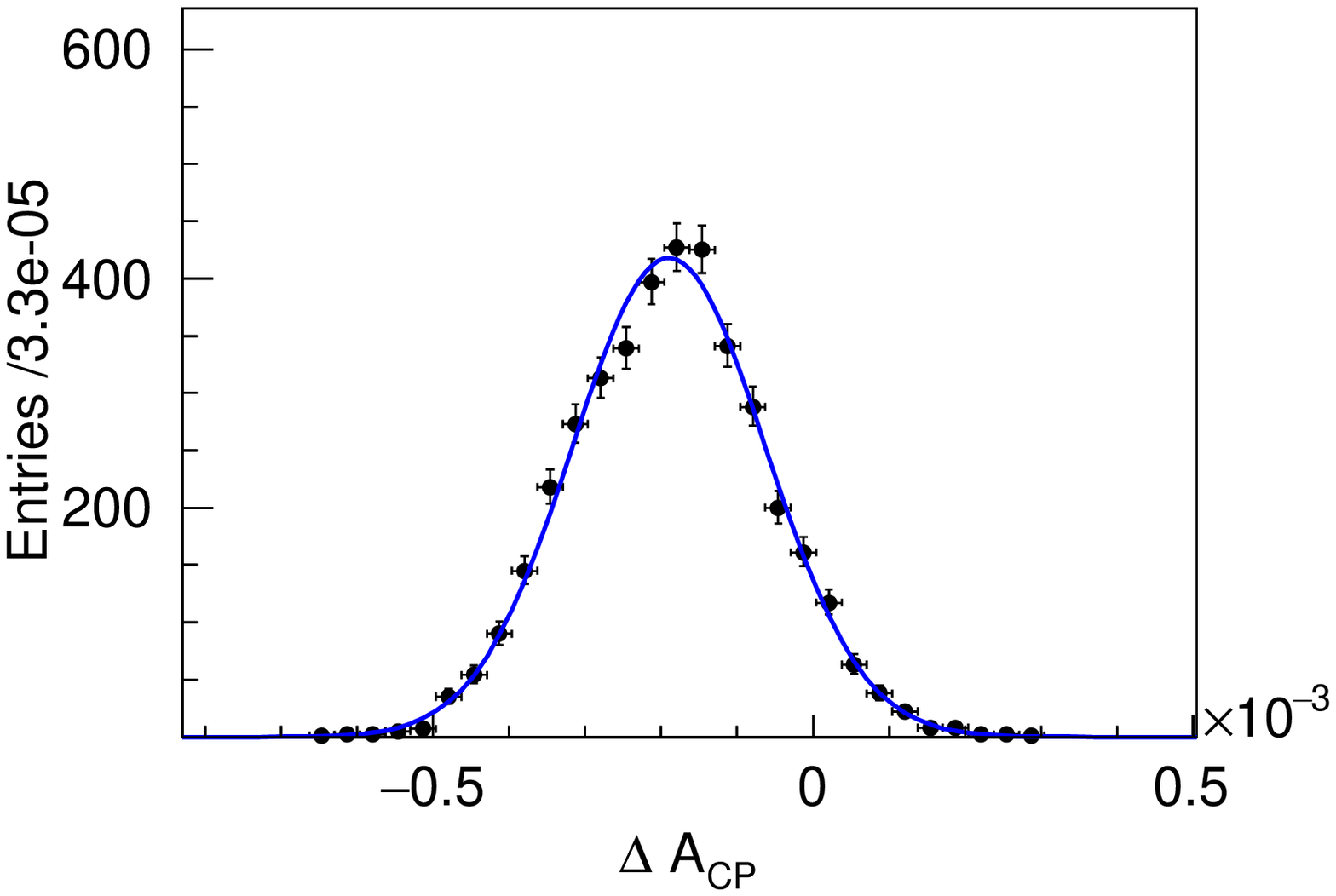}
    }
    \subfigure[]{
        \includegraphics[width=0.8\linewidth]{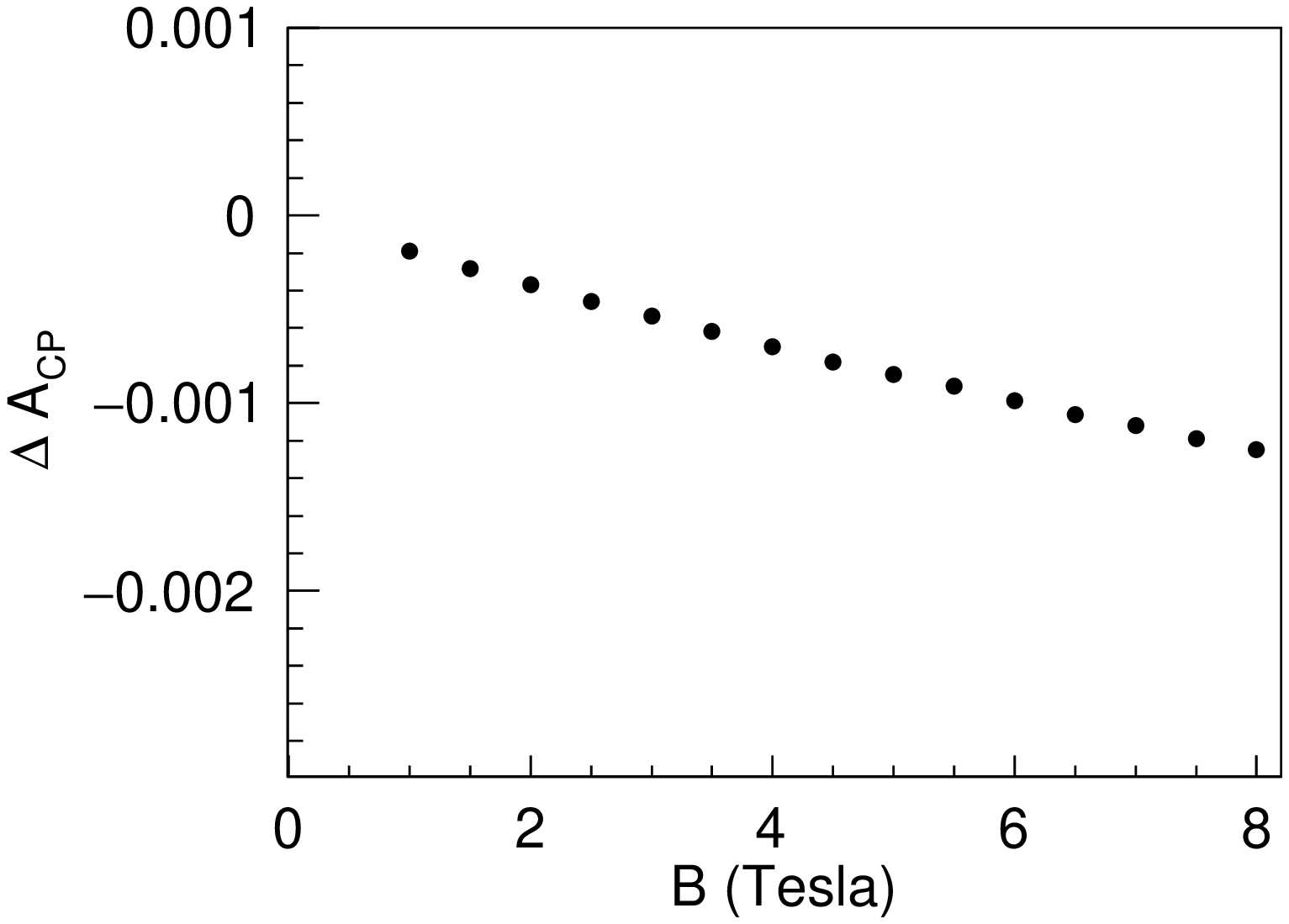} ~~~~~
    \label{fig:varyB}
    }
    \caption{(a)The distribution of $\Delta A_{CP}$ from 4000 times fits to the toy
    MC sample. The average of $\Delta A_{CP}$ and corresponding uncertainty are
    determined by fit to the distribution with a Gaussian function. 
    (b) The $\Delta A_{CP}$ will deviated more from zero, if the
    magnitude of the external magnetic field is increasing.}
    \label{fig:acpandvaryB}
\end{figure}
where $i$ denote the fit results from the $i$-th toy MC sample, $n$ the
number of total fit results, and take $\alpha_{-}^{\rm truth} +
\alpha_{+}^{\rm truth} =0$ because CP conservation is assumed. The size of
$\Delta A_{CP}$ is several times that of the $A_{CP}$ predicted by
the SM, as shown in Fig. \ref{fig:acp}. 
What's more, as we expect, the larger the magnitude of the magnitude
field $B$ is, the farther off zero the corresponding $\Delta A_{CP}$ will be,
as shown in Fig. \ref{fig:varyB}.
The values of $\alpha_{\psi}$ and $\Phi$ will also be derived from the truth
values relatively about 0.07\% and 0.01\%, respectively when the spin precession
neglected in the experiment.

\section{Summary}

In this work, we consider the spin precession of hyperon in the magnetic field
of the detector and give the differential cross-section for the global decay
chain. The corrected term is proportional to the lifetime of $\Lambda$ and the magnetic field.
The effects of spin precession are also estimated, based on the MC simulation.
The polarization of $\Lambda$ will be changed.
We find that a deviation of order $10^{-5}$ on the CP asymmetry will be
induced once neglecting the spin precession, which is the same level as that
from the SM prediction.
As well, a small deviation of $\alpha_{\psi}$ and $\Phi$ will be caused due to this effect.
The effect of the Larmor precession of hyperon in the external  magnetic field
has been also studied in Refs.~\cite{Kharzeev:2015znc,Guo:2019joy,Deng:2018frf}.
Following the method in this work, the effect could be easily extended to other
hyperon pair production at BESIII, such as $\Xi\bar{\Xi}$, $\Sigma^{0}
\bar{\Sigma}^{0}$, $\Omega\bar{\Omega}$, etc. In the further, the
super-tau-charm factor will reach a sensitivity  of $10^{-4}$ or even $10^{-5}$
\cite{Bigi:2017eni}, 
we suggest that one should consider the effect due to spin precession of
hyperons, so that one can determine the value of
$\alpha_{\pm}$ and CP asymmetry correctly in the experiment.

\begin{acknowledgments}
    The authors especially thank  H.-n. Li, Mao-Zhi
    Yang and Xian-Wei Kang for useful discussions.  This work is supported in part by the National Natural
    Science Foundation of China under Contracts Nos.~11335009,
    11125525, 11675137, 11875054, 11935018, the Joint Large-Scale Scientific Facility Funds of the NSFC and CAS
    under Contract No.~U1532257, CAS under Contract No.~QYZDJ-SSW-SLH003, and the National Key Basic Research Program of China under Contract
    No.~2015CB856700.
\end{acknowledgments}

\end{document}